\documentclass[aps,prl,secnumarabic,twocolumn,groupedaddress,showpacs]{revtex4}
\usepackage{amssymb}

\usepackage{amsmath}

\input{tcilatex}

\begin{document}

\title{Anomalous Hall effect for the phonon heat conductivity in a
paramagnetic dielectric}
\author{Y. Kagan}
\email{kagan@kiae.ru}
\author{L. A. Maksimov}
\affiliation{Kurchatov Institute, Moscow 123182, Russia}
\date{\today}

\begin{abstract}
The theory of anomalous Hall effect for the heat transfer in a paramagnetic
dielectric, discovered experimentally in \cite{1}, is developed. The
appearance of the phonon heat flux normal to both the temperature gradient
and the magnetic field is connected with the interaction of magnetic ions
with the crystal field oscillations. In crystals with an arbitrary phonon
spectrum this interaction creates the elliptical polarization of phonons.
The kinetics related to phonon scattering induced by the spin-phonon
interaction determines an origin of the off-diagonal phonon density matrix.
The combination of the both factors is decisive for the phenomenon under
consideration.
\end{abstract}

\pacs{66.70.+f, 72.15.Gd, 72.20.Pa}
\maketitle

\bigskip

\bigskip

\bigskip

\bigskip

\bigskip

\bigskip

\bigskip

\bigskip

\bigskip

\bigskip

% Force line breaks with \\


%\affiliation{ Second institution and/or address\\



A novel interesting phenomenon is found experimentally in recent paper \cite%
{1}. The matter concerns an analog of the anomalous Hall effect (AHE) for
the heat conductivity of an ionic paramagnetic dielectric. In fact, applying
magnetic field $\vec{B}$ in the direction normal to heat flow $\vec{ j}$,
the authors have discovered an appearance of the heat transfer in the
direction normal to $\vec{B}$ and $\vec{j}$. For the complete lack of free
charged carriers and negligible role of the spin-spin coupling at the
parameters concerned, the transverse flow is naturally associated with the
evolution of the phonon system. However, the magnetic field does not
directly act on phonons and only polarizes paramagnetic ions. It is the
coupling between phonons and subsystem of isolated ions carrying magnetic
moment $\vec{M}$ that determines the formation of the picture observed. Note
in this aspect that the phenomenon concerned is an analog of the AHE in the
paramagnetic phase of a ferromagnetic above the Curie point, discovered a
long time ago \cite{2}. An existence of the effect found in \cite{1} has
been confirmed in work \cite{3}. In both \cite{1} and \cite{3} the choice of
terbium gallium garnet (TbGG) was not accidental. In this compound for $%
T<10K $ the heat conductivity due to spin-phonon coupling (SPC) proves to be
about two orders of the magnitude smaller as compared with gadolinium garnet %
\cite{4} of the identical structure.

\qquad As it is known, in most of ion dielectrics the spin-phonon relaxation
is determined by the two-phonon processes, see, e.g., \cite{5} and \cite{6}.
Usually this supposes the presence of the Kramers doublet in the ground
state and the transitions via virtual excitation of higher Kramers doublets.
Trivalent ion $Tb^{3+}$ has the even number of f-electrons and the Kramers
degeneration is absent. However, for rare earth ions with the even number of
f-electrons, an appearance of the quasi-doublet structure is typical for
splitting the multiplet in the crystalline field with the level spacing $%
\varepsilon _{12}$ of about several Kelvins in the ground state. TbGG is a
striking example of the similar compound, see, e.g., \cite{7}. From the
general symmetry reasons and the direct analysis \cite{8,9,10} one can
conclude that the SPC under conditions concerned reduces to a scalar
production of magnetic moment $\vec{M}$ and the orbital moment related to
the motion of surrounding ions. At low temperatures when the long wave
acoustic branches alone are excited, the displacement of all atoms in the
elementary cell for each mode is the same. One can show in this case that
the total orbital moment of atoms in the elementary cell reduces to the
angular moment of the cell center of gravity. We will assume that the
magnetic ordering temperature $T_{c}\ll T$ and the magnitude of magnetic
field allows us to neglect the dipole-dipole interaction between ions. This
gives a possibility to analyze the phenomenon adequately, considering a
simple model system with one magnetic atom in the elementary cell and
assuming high symmetry of SPC. Thus, the Hamiltonian of SPC can be
represented as the spin-orbit interaction (see, e.g., \cite{8}) 
\begin{equation}
H_{1}=g\sum_{n}\vec{s}_{n}(\vec{U}_{n}\times \vec{P}_{n}),.  \label{1}
\end{equation}%
where $\vec{U}_{n}$ and $\vec{P}_{n}$ are the vectors of displacement and
momentum of the center of gravity in $n$-th elementary cell and $\vec{s}_{n}$
is the isospin typical for the quasi-doublet. For the isotropic SPC, one has 
$\vec{s}_{n}\parallel \vec{M}_{n}.$ Assuming that the spacing from the
ground quasi-doublet to higher levels is $\Delta \varepsilon \gg T$,
magnetic moment $\vec{M}_{n}$ is determined by the parameters of the
quasi-doublet components with the regard of their occupation. In the
approximation linear in $H_{1}$ we can replace $\vec{M}_{n}$ and,
correspondingly $\vec{s}_{n}$, with the quantities $\vec{M}$ and $\vec{s}$
averaged over the crystal.

The present paper is devoted to determining the heat conduction tensor $%
\varkappa _{odd}^{ab}\left( \vec{B}\right) $ antisymmetric and odd in
magnetic field $\vec{B}$ [11]. The heat conduction problem is solved at low
temperatures when the heat is carried with long wave acoustic phonons. A
fraction of heat flow directed along $\left[ \vec{M}\times \nabla T\right] $
arises from the elliptic renormalization of the polarization vectors of
acoustic phonons with involving (\ref{1}). The renormalization of the sound
velocity does not take place. The effect occurs at an arbitrary character of
the phonon dispersion spectrum and, in fact, at an arbitrary mechanism of
phonon scattering and phonon-phonon coupling. In this communication we
restrict ourselves with considering phonon scattering at magnetic ions with
the involvement of the level system which appears in the crystal field. The
elements of density matrix $\rho _{ss^{\prime }}(\vec{k})=\left\langle
a_{ks}^{+}a_{ks^{\prime }}\right\rangle $ off-diagonal in branches and found
from the generalized Boltzmann kinetic equation play a key role in the
formation of $\varkappa _{odd}^{ab}.$

In recent paper \cite{12} there has been attempted to explain theoretically
the phenomenon observed in \cite{1}. Taking SPC in the form analogous to (%
\ref{1}), the authors have chosen a simplified model for the phonon spectrum
in which two coincident transverse modes always exist for an arbitrary wave
vector. Such degeneration is decisive for the consideration on the whole. In
fact, in the general case the acoustic branches are not degenerated in most
part of phase volume. The regions in which one can meet degeneration occupy,
as a rule, relatively small fraction of the phase volume. This means a
necessity of the kinetic consideration with involving the phonon scattering
mechanisms.

\qquad 2. Treating the phonon system in the harmonic approximation, we write
the general Hamiltonian of the system with (\ref{1}) as%
\begin{equation}
H=H_{0}+H_{1},\ H_{0}=\sum_{n}\frac{1}{2m_{0}}\vec{P}_{n}^{2}-\frac{m_{0}}{2}%
\sum_{nn^{\prime }}D_{nn^{\prime }}^{ab}U_{n}^{a}U_{n^{\prime }}^{b}.
\label{2}
\end{equation}%
Here $m_{0}$ is the total mass of the elementary cell. Let us determine the
Hamilton's equations from (\ref{2}). For velocity $V_{n}^{a}$, one has
straightforwardly 
\begin{equation}
V_{n}^{a}=\dot{U}_{n}^{a}=P_{n}^{a}/m_{0}+ge_{abc}s^{b}U_{n}^{c},  \label{3}
\end{equation}%
where $e_{abc}$ is the antisymmetric third-rank unit tensor. Using explicit
expression (\ref{3}), the equation of motion can be transformed to the form 
\begin{equation}
-\ddot{U}_{n}^{a}=\sum_{n^{\prime }}D_{nn^{\prime }}^{ab}\left(
U_{n}^{b}-U_{n^{\prime }}^{b}\right) +2ge_{abc}\dot{U}_{n}^{b}s^{c}.
\label{4}
\end{equation}

Assuming further the consideration in the linear approximation for the
spin-orbit interaction (SOI), we replaced $P_{n}^{b}$ with $m_{0}\dot{U}%
_{n}^{b}$ in the last term. A usual Fourier-transformation of linear
equation (\ref{4}) results in the dispersion equation%
\begin{equation}
\omega ^{2}U_{\vec{k}}^{a}=D_{\vec{k}}^{ab}U_{\vec{k}}^{b}-2ig\omega
e_{abc}U_{\vec{k}}^{b}s^{c},  \label{5}
\end{equation}%
where%
\begin{equation}
D_{\vec{k}}^{ab}=\sum_{\vec{R}}D_{\vec{R}}^{ab}(1-e^{i\vec{k}\vec{R}}),\
\sum_{\vec{R}}D_{\vec{R}}^{ab}=0.  \label{06}
\end{equation}%
Here $\vec{R}=\vec{r}_{n}-\vec{r}_{n^{\prime }},$ $\vec{r}_{n}$\ is a
position of the center of gravity in $n$-th cell. In zeroth order in SOI the
solution of a set of equations (\ref{5}) determines the dispersion law $%
\omega _{s}(\vec{k})$ for three acoustic branches and, correspondingly,
three orthonormal polarization vectors $e_{s}^{(0)a}(\vec{k})$ in the long
wave limit. The latter ones can be chosen as real. In the approximation
linear in the SOI one can conclude from (5) that the phonon spectrum remains
unchanged and the polarization vectors alone renormalize. Let us introduce
notation%
\begin{equation}
U_{s}^{a}(\vec{k})=e_{s}^{(0)a}(\vec{k})+\delta e_{s}^{a}(\vec{k}).
\label{07}
\end{equation}

Then from equation (\ref{5}) one finds the small rotation of the
polarization vector for $s$-th branch 
\begin{equation}
\delta \vec{e}_{1}=-2ig\omega _{1}[\frac{\left( \left( \vec{e}_{2}\times 
\vec{e}_{1}\right) \vec{s}\right) }{(\omega _{1}^{2}-\omega _{2}^{2})}\vec{e}%
_{2}+\frac{\left( \left( \vec{e}_{3}\times \vec{e}_{1}\right) \vec{s}\right) 
}{(\omega _{1}^{2}-\omega _{3}^{2})}\vec{e}_{3}].  \label{09}
\end{equation}%
Accordingly, for $s=2$ and $3$. In the linear approximation in SOI the
condition of orthonormality holds for as $e_{s}^{a\ast }e_{s^{\prime
}}^{a}=\delta _{ss^{\prime }}$.

To find the phonon energy flow $j^{c}$ with the presence of SOI, one can
employ the general results obtained in \cite{13}. The successive use of the
continuity equation, reflecting the local law of energy conservation for the
slow spatial temperature variation, results in the expression%
\begin{equation}
j^{c}=\frac{1}{2V}m_{0}\sum_{nn^{\prime }}R_{nn^{\prime }}^{c}D_{nn^{\prime
}}^{ab}U_{n}^{a}V_{n^{\prime }}^{b}.  \label{010}
\end{equation}

Hereafter the volume is V=1. In this expression the velocity $V_{n^{\prime
}}^{b}$\ is determined by relation (\ref{3}). Let us expand vectors $%
U_{n}^{a}$ (and $V_{n}^{a}=\dot{U}_{n}^{a})$ into the normal modes in the
representation of secondary quantization ($\hbar =1$)%
\begin{equation}
U_{n}^{a}=\sum_{ks}\sqrt{\frac{1}{2m_{0}N\omega _{ks}}}\exp \left( i\vec{k}%
\vec{r}_{n}\right) [e_{s}^{a}(\vec{k})a_{ks}+e_{s}^{a\ast }(-\vec{k}%
)a_{-ks}^{+}].  \label{011}
\end{equation}%
Substituting these expressions into (\ref{010}) and employing relation (\ref%
{06}), we find for the averaged operator of the energy flow%
\begin{equation}
\left\langle j^{c}\right\rangle =\frac{1}{4}\sum_{kss^{\prime }}\{(\sqrt{%
\frac{\omega _{ks}}{\omega _{ks^{\prime }}}}+\sqrt{\frac{\omega _{ks^{\prime
}}}{\omega _{ks}}})\left( \nabla _{k}^{c}D_{k}^{ab}\right) e_{ks}^{a\ast
}e_{ks^{\prime }}^{b}\}\rho _{ss^{\prime }}(\vec{k}),  \label{012}
\end{equation}%
where $\rho _{ss^{\prime }}(\vec{k})=\left\langle a_{ks}^{+}a_{ks^{\prime
}}\right\rangle $. Keeping the approximation linear in SOI, we neglect
anomalous averages $\left\langle a_{-ks}a_{ks^{\prime }}\right\rangle $ and $%
\left\langle a_{ks}^{+}a_{-ks^{\prime }}^{+}\right\rangle $ in (\ref{012}).
For zero order approximation in SOI the polarization vectors in (\ref{012})
are real and the expression in the figure brackets is symmetric over indices 
$s,s^{\prime }$. Thus only the symmetric\ and real component of density
matrix gives nonzero contribution to ((\ref{012})). Assuming $\rho
_{ss^{\prime }}(\vec{k})=n_{s}(\vec{k})\delta _{ss^{\prime }}$ and the
relation%
\begin{equation}
\left( \nabla _{k}^{c}D_{k}^{ab}\right) e_{ks}^{a}e_{ks}^{b}=\partial \omega
_{ks}^{2}/\partial k^{c}=2\omega _{ks}C_{ks}^{c},  \label{014}
\end{equation}%
where $\vec{C}_{s}=\partial \omega _{s}/\partial \vec{k}$ is the sound
velocity, from (\ref{012}) we find an ordinary expression for the heat flow 
\begin{equation}
\left\langle j_{0}^{c}\right\rangle =\sum_{ks}\omega _{ks}C_{ks}^{c}n_{s}(%
\vec{k})  \label{015}
\end{equation}

The contribution linear in SOI appears in (\ref{012}) with involving the
renormalization of the polarization vectors (\ref{07}) 
\begin{equation}
e_{ks}^{a\ast }e_{ks^{\prime }}^{b}\rightarrow -\left( \delta
e_{ks}^{a}\right) e_{ks^{\prime }}^{(0)b}+e_{ks}^{(0)a}\left( \delta
e_{ks^{\prime }}^{b}\right)  \label{017}
\end{equation}%
Substituting relations (\ref{09}) into this expression, we find that the
contribution diagonal in modes vanishes. The off-diagonal term with $ks=1$
and $ks^{\prime }=2$ equals 
\begin{equation}
\begin{array}{c}
2ig\frac{\left( \vec{e}_{2}\times \vec{e}_{1}\right) \vec{s}}{(\omega
_{1}^{2}-\omega _{2}^{2})}\left( \omega _{1}e_{2}^{a}e_{2}^{b}-\omega
_{2}e_{1}^{a}e_{1}^{b}\right) \\ 
-2ig\frac{\left( \vec{e}_{3}\times \vec{e}_{2}\right) \vec{s}}{(\omega
_{2}^{2}-\omega _{3}^{2})}\omega _{2}e_{1}^{a}e_{3}^{b}+2ig\frac{\left( \vec{%
e}_{3}\times \vec{e}_{1}\right) \vec{s}}{(\omega _{1}^{2}-\omega _{3}^{2})}%
\omega _{1}e_{3}^{a}e_{2}^{b}%
\end{array}
\label{016}
\end{equation}%
With substitution (\ref{017}),(\ref{016}) into (\ref{012}) the expression in
the figure brackets, proves to be antisymmetric over indices 1 and 2. The
contribution linear in SOI in (\ref{012}) does not vanish if the
off-diagonal terms of density matrix are nonzero. In the expression (\ref%
{016}) the first term plays a key role. To avoid the cumbersome expressions,
we will find the approximate value for the flow (\ref{012}), keeping the
first term alone $\ \left\langle j_{SO}^{a}\right\rangle =\left\langle
j_{12}^{a}\right\rangle +\left\langle j_{23}^{a}\right\rangle +\left\langle
j_{31}^{a}\right\rangle $, where 
\begin{equation}
\begin{array}{c}
\left\langle j_{12}^{a}\right\rangle =ig\sum_{k}\left( \sqrt{\frac{\omega
_{1}}{\omega _{2}}}+\sqrt{\frac{\omega _{2}}{\omega _{1}}}\right) \\ 
\frac{2\omega _{1}\omega _{2}\left( \vec{e}_{2}\times \vec{e}_{1}\right) 
\vec{s}}{(\omega _{1}^{2}-\omega _{2}^{2})}\left( C_{2}^{a}-C_{1}^{a}\right)
\rho _{12}.%
\end{array}
\label{8}
\end{equation}%
\qquad

To find the flow finally, it is necessary to determine the imaginary part of
nonequilibrium off-diagonal density matrix $\rho _{12}$. We employ a
conventional procedure to derive the kinetic equation, see, e.g., \cite{14}.
The general equation of evolution reads 
\begin{equation}
-i\partial _{t}\rho _{kk^{\prime }}=\left\langle [a_{k}^{+}a_{k^{\prime
}},H_{0}+H^{\prime }]\right\rangle  \label{19}
\end{equation}%
Here $H^{\prime }$ is the Hamiltonian for the phonon scattering and $H_{0}$
is the starting Hamiltonian of noninteracting phonons (\ref{2}). To simplify
notations, we introduce a generalized index for the mode $k\equiv \vec{k},s$%
. In the frequency representation%
\begin{equation}
\left( \omega +\omega _{k}-\omega _{k^{\prime }}\right) \rho _{kk^{\prime
}}=I_{kk^{\prime }},\ I_{kk^{\prime }}=\left\langle [a_{k}^{+}a_{k^{\prime
}},H^{\prime }]\right\rangle  \label{20}
\end{equation}
The formal stationary solution $\left( \omega _{k}\neq \omega _{k^{\prime
}}\right)$ reads 
\begin{equation}
\rho _{kk^{\prime }}\mid _{\omega \rightarrow 0}=\frac{I_{kk^{\prime }}}{%
\omega _{k}-\omega _{k^{\prime }}}.  \label{21}
\end{equation}

At low temperatures, as we have already mentioned, the main mechanism of
scattering is connected with the coupling between phonons and magnetic ions.
The splitting of rare-earth ion multiplet with the crystalline field $V_{cr}$
and the reduction of the symmetry for oscillating part $V_{cr}$ make
decisive the Raman two-phonon scattering via virtual excitation of the other
multiplet levels \cite{5,6}. For garnets like $Tb^{3+}$, the ground state is
a quasi-doublet with the small spacing of about several Kelvins between the
doublet components.

Considering the region of low temperatures $T\leqslant 6K$, we assume that
the kinetics is determined by the scattering of phonons at this
quasi-doublet. The higher levels lie at $\varepsilon _{i}\geq 50K$ and we
neglect rescattering through them. In this case Hamiltonian $H^{\prime }$
can be represented in the general form%
\begin{equation}
H^{\prime }=\frac{1}{N}\sum_{\vec{r}_{n}}\sum_{fg\alpha }e^{i(\vec{g}-\vec{f}%
)\vec{r}_{n}}A_{fg}^{\alpha \alpha }\xi _{n}^{\alpha }a_{f}^{+}a_{g}.
\label{22}
\end{equation}
Here $\alpha =1,2$ numerates the doublet levels and $\xi _{n}^{\alpha }=1,0$%
. The commutator in (\ref{20}) equals 
\begin{equation*}
\left\langle \lbrack a_{k}^{+}a_{k^{\prime }},\xi _{n}^{\alpha
}a_{f}^{+}a_{g}]\right\rangle =\left\langle a_{k}^{+}\xi _{n}^{\alpha
}a_{g}\right\rangle \delta _{k^{\prime }f}-\left\langle a_{f}^{+}\xi
_{n}^{\alpha }a_{k^{\prime }}]\right\rangle \delta _{kg}
\end{equation*}
Using equations (\ref{20}) for these correlators, we obtain some quadruple
operator correlators.

Restricting with the Born approximation, we decouple these correlators in
the mean-field approximation. As a result, we have at $\omega \rightarrow 0$ 
\begin{equation}
\begin{array}{c}
I_{kk^{\prime }}=-i\pi \frac{1}{N}\sum_{g\alpha }P_{\alpha }(1-P_{\alpha
})A_{k^{\prime }g}^{\alpha \alpha }A_{gk}^{\alpha \alpha } \\ 
\{\delta \left( \omega _{k}-\omega _{g}\right) (n_{k}-n_{g})+\delta \left(
\omega _{k^{\prime }}-\omega _{g}\right) \left( n_{k^{\prime }}-n_{g}\right)
\},%
\end{array}
\label{023}
\end{equation}%
where $P_{\alpha }=\left\langle \xi _{n}^{\alpha }\right\rangle $ is the
occupation value of level $\alpha $. From (\ref{023}) we can conclude that $%
I_{kk^{\prime }}=I_{k^{\prime }k}$. As a result, we see that off-diagonal
matrix $\mathbf{\rho }_{kk^{\prime }}$ (\ref{21}) is antisymmetric. For the
equilibrium distribution, $I_{kk^{\prime }}=0$. Nonzero result appears only
for nonequilibrium distribution $f_{k}=n_{k}-N_{k}^{(0)}$ due to temperature
gradient. In the $\tau $-approximation the expression (\ref{023}) can be
represented as%
\begin{equation}
I_{kk^{\prime }}=-i\frac{1}{2}\left( \Omega _{kk^{\prime }}f_{k}+\Omega
_{k^{\prime }k}f_{k^{\prime }}\right) ,  \label{24}
\end{equation}%
\begin{equation}
\Omega _{kk^{\prime }}=2\pi \frac{1}{N}\sum_{g\alpha }P_{\alpha }\left(
1-P_{\alpha }\right) \left( A_{k^{\prime }g}^{\alpha \alpha }A_{gk}^{\alpha
\alpha }\right) \delta \left( \omega _{k}-\omega _{g}\right).  \label{241}
\end{equation}%
The usual solution of the problem for the longitudinal heat conduction yields%
\begin{equation}
f_{k}=-\left( \omega _{k}/\Omega _{kk}T^{2}\right) N_{k}^{(0)}\left(
1+N_{k}^{(0)}\right) \vec{C}_{k}\nabla T  \label{26}
\end{equation}

The substitution of the off-diagonal density matrix (\ref{21}) into (\ref{8}%
) with involving (\ref{24})-(\ref{26}) solves the problem for the heat
transfer due to SOI. The relaxation frequencies (\ref{241}) depend on a
product of the transition amplitudes with $k\neq k^{\prime }$ instead of a
usual square of the modulus of the transition amplitude when everything is
determined by the diagonal density matrix. This typical feature is inherent
in any scattering mechanism, in particular, due to phonon anharmonicity.

Let us return to the notations adopted in (\ref{8}) and make a significant
remark about the symmetry for a product of polarization vectors with the
same wave vector $\vec{k}$ entering this expression. In the long wave
approximation for the crystals of the sufficiently high symmetry the
off-diagonal elements of the dynamical matrix read $D_{k}^{ab}\sim
k^{a}k^{b} $, e.g., \cite{15}. Rewrite the dispersion equation (\ref{5}) for
zero order approximation in SOI in the form $\left( \omega
_{ks}^{2}-D_{k}^{aa}\right) e_{ks}^{a}=\sum_{b\neq a}D_{k}^{ab}e_{ks}^{b}$.
One can show that, with varying the sign of the separate projection of wave
vector, $D_{k}^{aa}$ and $\omega _{ks}^{2}$ remain unchanged. Then from the
given equation it follows that the projection of polarization vector $%
e_{ks}^{a}$ changes its sign for the substitution $k^{a}$ with $\left(
-k^{a}\right) $ conserving the sign of the other projections. This property
can be expressed in terms%
\begin{equation}
e_{ks}^{a}=\tilde{e}_{s}^{a}\left( \vec{k}\right) signk^{\alpha }.
\label{27}
\end{equation}%
Here $\tilde{e}_{s}^{a}\left( \vec{k}\right) $ is a unit vector remaining
unchanged with varying the sign of an arbitrary projection of wave vector $%
\vec{k}$.

Let magnetic field and magnetic moment $\vec{M}$ be in the direction of the $%
z$-axis, while the longitudinal $\nabla T$ be in the direction of the $x$%
-axis. Then, substituting (\ref{21}) and (\ref{24})-(\ref{27}) into (\ref{8}%
), we find, restoring $\hbar$, $k_{B}$ and volume $V$, for the transverse
component of the heat conduction tensor $\varkappa ^{yx}=\varkappa
_{12}^{yx}+\varkappa _{23}^{yx}+\varkappa _{31}^{yx}$, 
\begin{equation}
\begin{array}{c}
\varkappa _{12}^{yx}\simeq k_{B}\frac{g\hbar ^{2}}{V}\sum_{k}s^{z}\left( 
\tilde{e}_{1}\times \tilde{e}_{2}\right) ^{z}\left( \sqrt{\frac{\omega _{1}}{%
\omega _{2}}}+\sqrt{\frac{\omega _{2}}{\omega _{1}}}\right) \\ 
\times \frac{2\omega _{1}\omega _{2}signk^{x}signk^{y}}{\left( \omega
_{1}^{2}-\omega _{2}^{2}\right) \left( \omega _{1}-\omega _{2}\right) }%
\left( c_{1}^{y}-c_{2}^{y}\right) \left(
F_{12}c_{1}^{x}+F_{21}c_{2}^{x}\right) , \\ 
F_{12}=-\frac{1}{2}\omega _{1}\frac{\Omega _{12}}{\Omega _{11}}\frac{1}{%
\left( k_{B}T\right) ^{2}}N_{1}^{(0)}\left( 1+N_{1}^{(0)}\right)%
\end{array}
\label{28}
\end{equation}

Recall that \ $s^{z}\equiv \left\langle s^{z}\right\rangle $. The expression
obtained demonstrates an existence of the anomalous phonon Hall effect under
comparatively general conditions conserving the intrinsic symmetry. It is
interesting that the kinetics, reflecting phonon scattering character,
enters (\ref{28}) only as a ratio like $\Omega _{12}/\Omega _{11}$. This
holds for any dominant mechanism of scattering. Under these conditions the
transverse heat transfer is determined mainly by the spectral properties of
the phonon system. Note, as it follows from (\ref{28}), that a quantitative
enhancement of the effect takes place if the region in which branches $%
\omega _{ks}$ come close is noticeable. We will give a quantitative estimate
of the effect, assuming that the role of these regions in the integral
determining (\ref{28}) is limited in the general case.

For the estimate, one should determine the magnitude of parameter $g$
introduced formally. The SOI (\ref{1}) due to two-phonon processes
accompanying vibrations of the crystalline field has the same origin as
Hamiltonian $H^{\prime }$ (\ref{22}). The Hamiltonian of single-phonon
interaction is written usually for a mode $\left( \vec{k}s\right) $ in the
simplified form $V^{\prime }\left( U_{ks}/a\right) (ka)$, where $V^{\prime }$
is close to the magnitude of the static crystalline field (see, e.g., \cite%
{6}). Factor $(ka)$ appears with regard to the relative motion of magnetic
ion and surrounding atoms. In second order in this interaction, the standard
unitary transformation results in the spin-orbit interaction as (see \cite{8}%
)

\begin{equation*}
\left\langle s^{z}\right\rangle V^{\prime 2}U_{ks}V_{ks^{\prime }}\frac{%
\hbar }{\bar{C}^{2}}\frac{\omega _{ks}^{2}}{\varepsilon _{12}^{2}-\hbar
^{2}\omega _{ks}^{2}}
\end{equation*}

To simplify, we introduce here an averaged value for sound velocity $\bar{C}$
and take $\omega _{ks}\simeq \omega _{ks^{\prime }}$. Choosing two-mode
contribution to SOI, written in the form (\ref{1}), we have $\left\langle
s^{z}\right\rangle gm_{0}U_{ks}V_{ks^{\prime }}.$ In fact, we should mean
the displacement in $H_{1}$ as a relative displacement of ions. Thus in this
expression, as well as in the final one (\ref{28}), $\tilde{g}=g\left(
ka\right) ^{2}$ should be implied instead of $g$. Comparing the expressions
presented and assuming the temperature $T>\varepsilon _{12}$, we find $%
\tilde{g}\simeq V^{\prime 2}/\left( \hbar m_{0}\bar{C}^{2}\right) $.

Returning to (\ref{28}), let us make an integral dimensionless. Dividing
sound velocity by $\bar{C}$, we obtain $\varkappa _{12}^{yx}\sim 3\pi
^{-2}\left( \hbar \bar{C}/T\right) ^{2}\int dkk^{2}\ldots $ The remaining
part of the integrand is dimensionless. Putting that $\mu _{eff}B<k_{B}T$,
we find within the accuracy of a numerical factor%
\begin{equation}
\varkappa ^{yx}\simeq k_{B}\tilde{g}\left( k_{B}\Theta _{D}/\hbar \bar{C}%
\right) \left( T/\Theta _{D}\right) \left( \mu _{eff}B/k_{B}T\right) .
\end{equation}

Taking $V^{\prime }\simeq 50K$, $\bar{C}\simeq 4\ast 10^{5}cm/s$, $\Theta
_{D}\simeq 400K$, we arrive at $\varkappa ^{yx}\sim 10^{-7}\left( \mu
_{eff}B/k_{B}T\right) \left( W/cmK\right) $. Using the experimental value
for $\varkappa ^{xx}$ [1,4] at $\mu _{eff}B/k_{B}T\sim 1$, we find for the
Hall angle%
\begin{equation}
\eta =\varkappa ^{yx}/\varkappa ^{xx}\sim 2\ast 10^{-5}.  \label{30}
\end{equation}

In reality, this ratio may be larger since the numerical coefficient omitted
in (\ref{30}) can take the value significantly larger than unity due to the
character in the behavior of integrand (\ref{28}).

So, in the general case the anomalous phonon Hall effect originates from a
combination of two important factors. The first of them is associated with
an appearance of elliptic renormalization for phonon modes as a result of
the spin-orbit interaction between phonons and paramagnetic ions. The second
factor is connected with the proof that the off-diagonal density matrix of
phonons does not vanish at the kinetics of the longitudinal heat transfer
with the presence of temperature gradient $\nabla T$. The latter
predetermines a necessity to solve nonconventional kinetic equation in the
problem of phonon heat transfer. The model of a crystal and interaction used
in the paper is undoubtedly simplified. That is why, the fact that the
estimate obtained for the Hall angle (\ref{30}) with using the real
parameters proves to be comparable with the experimental magnitudes \cite%
{1,3} lying within interval $10^{-4}\div 10^{-5}$ seems very encouraging.

\end{document}